\pdfoutput=1
\documentclass[aps,prd,reprint,superscriptaddress]{revtex4-1}

\usepackage{graphicx}
\usepackage{amsmath}
\usepackage{amssymb}
\usepackage{amsfonts}
\usepackage{dcolumn}
\usepackage{dsfont}
\usepackage{latexsym}
\usepackage{rotating}
\usepackage{color}
\usepackage{latexsym}
\usepackage{bbm}
\usepackage{subfigure}
\usepackage{float}
\usepackage{epsfig}
\usepackage{psfrag}
\usepackage{natbib, hyperref}
\usepackage{bm}
\usepackage{amsthm}
\usepackage{eucal}
\usepackage{mathrsfs}
\usepackage{url}
\usepackage{braket}
\usepackage{ulem}
\usepackage{calligra}
\usepackage[T1]{fontenc}
\usepackage[utf8]{inputenc}
\usepackage{newunicodechar}
\usepackage{marvosym}
\usepackage[absolute]{textpos}
\usepackage[cal=cm]{mathalfa}
\usepackage{soul}
\usepackage{tcolorbox}

\usepackage{color}

\hypersetup{
colorlinks=true,final=true,
        linkcolor=blue,
        citecolor=blue,
        filecolor=blue,
        urlcolor=blue,
}

\begin{document}
\setlength{\abovedisplayskip}{3pt}
\setlength{\belowdisplayskip}{3pt}

% Change Figure prefix from FIG. to Fig. in bold
\setcounter{figure}{0}
\renewcommand{\figurename}{{\bf Fig.}}
\renewcommand{\thefigure}{{\bf \arabic{figure}}}
%=========================================================

\title{Interfacial phase frustration stabilizes unconventional skyrmion crystals}

\author{Narayan Mohanta}
\thanks{Email: narayan.mohanta@ph.iitr.ac.in}
\affiliation{Materials Science and Technology Division, Oak Ridge National Laboratory, Oak Ridge, TN 37831, USA}
\affiliation{Department of Physics and Astronomy, The University of Tennessee, Knoxville, TN 37996, USA}
\affiliation{Department of Physics, Indian Institute of Technology Roorkee, Roorkee 247667, India}

\author{Elbio Dagotto}
\affiliation{Materials Science and Technology Division, Oak Ridge National Laboratory, Oak Ridge, TN 37831, USA}
\affiliation{Department of Physics and Astronomy, The University of Tennessee, Knoxville, TN 37996, USA}

\begin{abstract} 
Chiral magnetic phases with an unconventional topological twist in the magnetization are of huge interest due to their potential in spintronics applications. Here, we present a general method to induce such exotic magnetic phases using interfacial phase frustration within artificially grown superlattices. To demonstrate our method, we consider a multilayer with two different chiral magnetic phases as the competing orders at the top and bottom and show, using Monte Carlo calculations, that the interfacial phase frustration is realized at the central layer. In particular, we obtain three unconventional phases: a checkerboard skyrmion crystal, an incommensurate skyrmion stripe, and a ferrimagnetic skyrmion crystal. In these frustration-induced phases, the spin chirality driven topological Hall conductivity can be largely enhanced. This method provides a playground to realize unconventional magnetic phases in any family of materials that can be grown in superlattices.
\end{abstract}

\maketitle

\noindent \small{\bf \\INTRODUCTION\\}
Magnetic frustration arises from the incompatibility of the interactions between magnetic degrees of freedom in a lattice with the underlying crystal geometry and often produces exotic ground states, such as spin ice and spin liquids~\cite{Bramwell_Science2001, Balents2010}. These frustration-stabilized magnetic phases opened a window to understand the fundamentals of magnetism and to use in technological applications, as exemplified by the search for quantum spin liquids~\cite{Kitaev2003,Savary_RPP2016,Batista_RPP2016}. The collective behavior of the spins, influenced by the frustration, often leads to chiral correlations in otherwise non-chiral ordered magnetic ground states~\cite{Taguchi_Science2001,Grohol_NatMat2005}.
Specifically, chiral magnetic textures are the focus of intense investigation because of their potential in spintronic applications, and for their role in producing topological Hall effects from the emergent magnetic field created by the magnetic texture~\cite{Schulz_NatPhys2012,Nagaosa_NatNano2013,Fert_NarRevMat2017}. 

A prominent example of such chiral magnetic texture with a continuous winding of magnetization is a skyrmion, stabilized commonly as a triangular crystal in various bulk compounds and interfaces by competing Heisenberg ferromagnetic exchange and relativistic Dzyaloshinskii-Moriya interaction (DMI)~\cite{Rossler_Nature2006, Muhlbauer_Science2009, Yu_Nature2010, Balents_PRL2014, Das_Nature2019, Mohanta_PRB2019}.  The skyrmion crystal phase was also theoretically predicted to be stabilized without DMI in geometrically-frustrated magnets or in highly-symmetric lattices with long-range Ruderman-Kittel-Kasuya-Yosida interaction~\cite{Okubo_PRL2012,Leonov_NatComm2015,Ozawa_PRL2017,Wang_PRL2020}. Three dimensional variants of skyrmions, \textit{viz.}, skyrmion strings~\cite{Park_NatNano2014}, chiral bobbers~\cite{Zheng2018} and hopfions~\cite{Kent_NatCommun2021} have been experimentally observed. However, the variety of these chiral magnetic textures beyond the skyrmion crystal is severely limited by the available magnetic interactions in bulk compounds.

%---------------------------------------------
\begin{figure}[b]
\begin{center}
\vspace{-0mm}
\epsfig{file=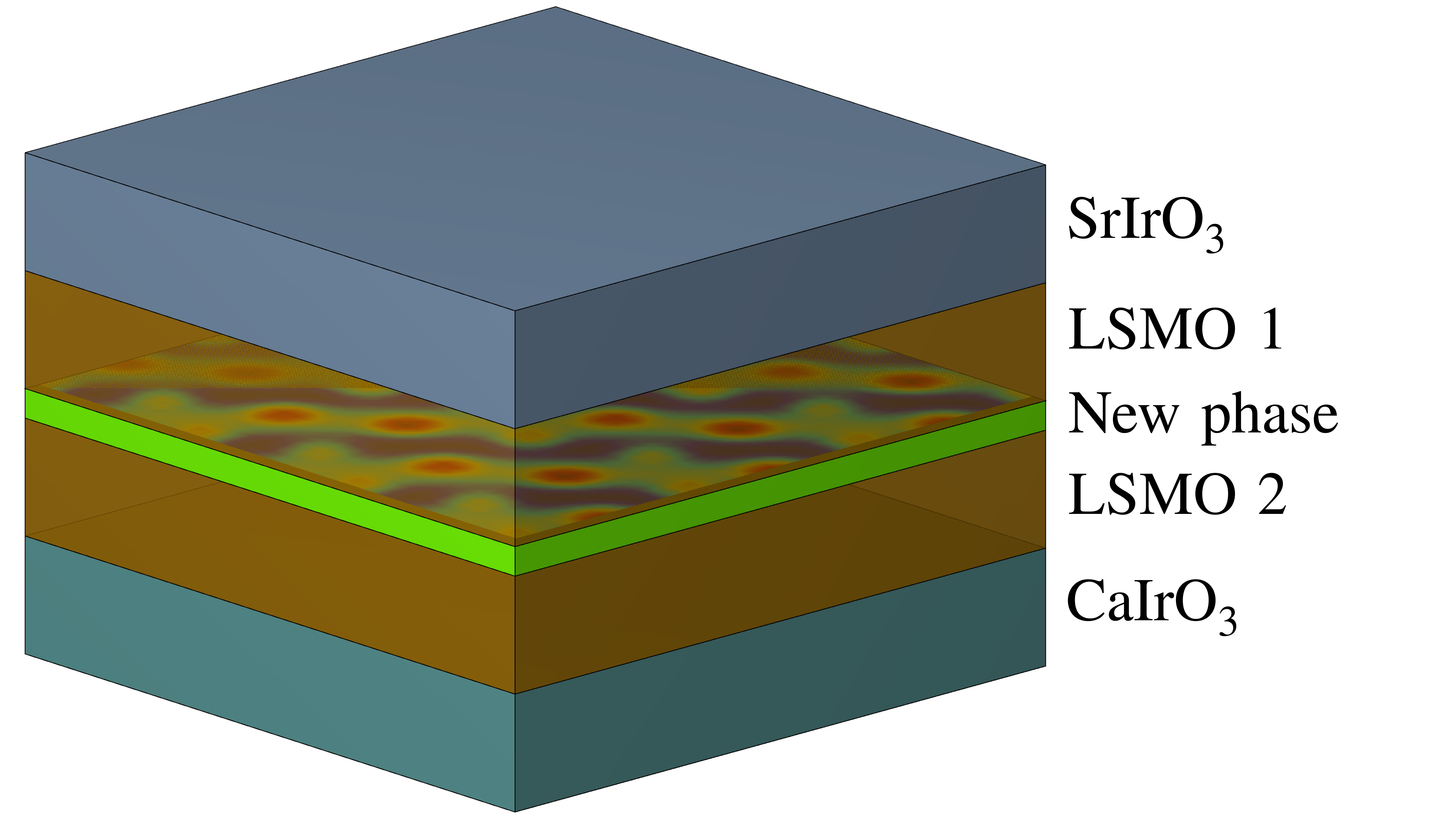,trim=0.0in 0.0in 0.0in 0.0in,clip=false, width=74mm}
\caption{{\bf Schematic illustration of a multilayer to realize ``interfacial phase frustration'.} Shown is a magnetic multilayer, involving different $5d$ compounds, top and bottom, and LSMO layers. An emergent phase will appear in the middle layer (green), between two LSMO layers each influenced by two different $5d$ compounds with different DMI strengths or different spiral orientations. The central ``green'' layer is {\it frustrated} by competing tendencies above and below.} 
\label{schematic}
\vspace{-2mm}
\end{center}
\end{figure}
%---------------------------------------------

%---------------------------------------------
\begin{figure*}[t]
\begin{center}
\epsfig{file=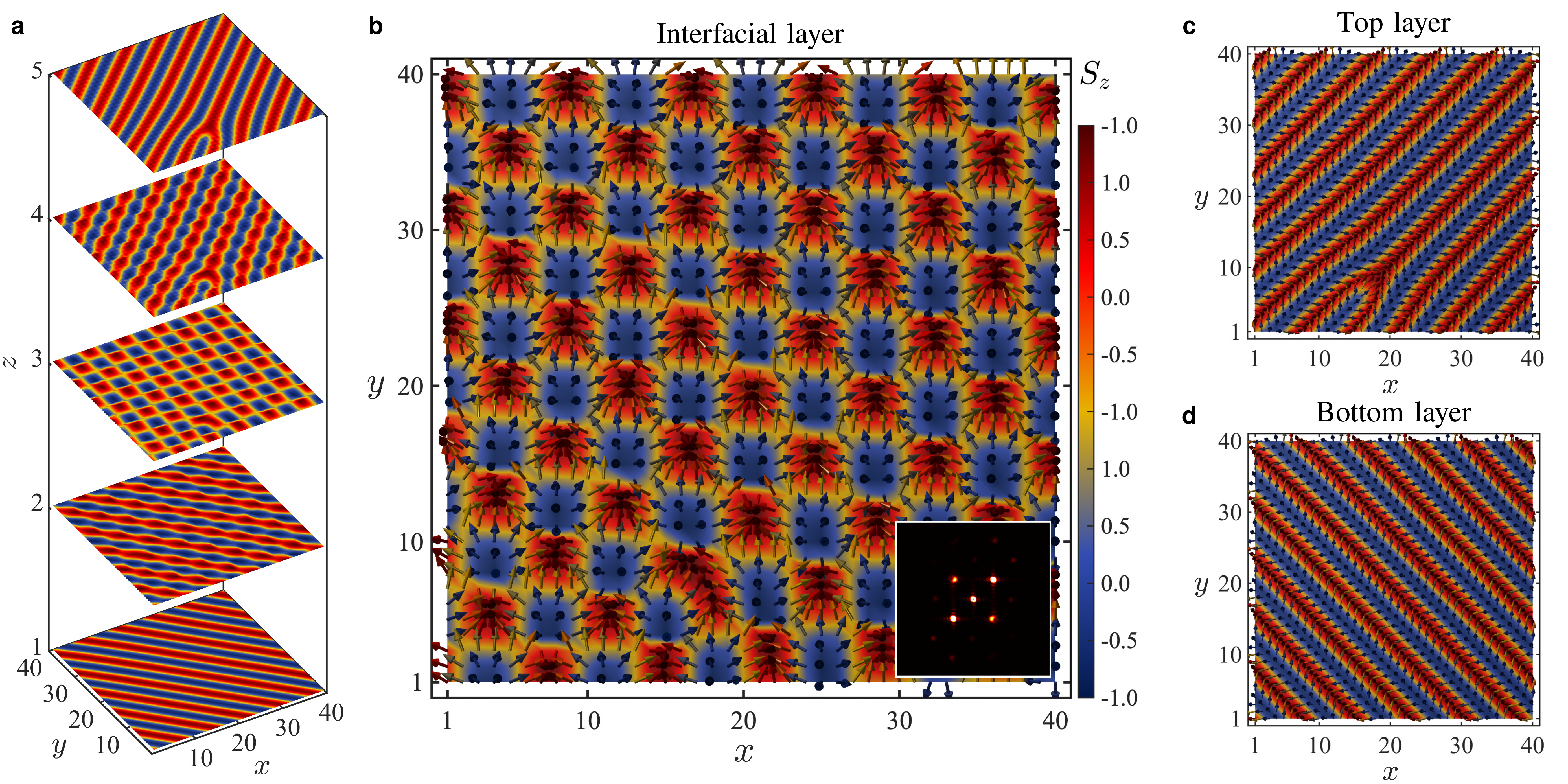,trim=0.0in 0.0in 0.0in 0.0in,clip=false, width=180mm}
\caption{{\bf Checkerboard skyrmion crystal.} {\bf a} Plot of the $z$-component of the spin $S_z$, obtained via MC annealing, in a five-layer heterostructure. {\bf b} Spin configuration of the frustration-stabilized checkerboard skyrmion crystal (CSkX) at the central layer ($z\!=\!3$). {\bf c} and {\bf d} are the canonical spin spirals at the top layer ($z\!=\!5$) and the bottom layer ($z\!=\!1$), oriented along mutually-orthogonal diagonals. Inset in {\bf b} shows the central layer spin structure factor $S(\mathbf{q})$ within the first Brillouin zone, revealing a double-\textit{q} magnetic order. The parameters used are $D\!=\!J$, $A\!=\!0$, $J\!=\!1$ (both top and bottom layers).}
\label{cskx}
\vspace{-2mm}
\end{center}
\end{figure*}
%---------------------------------------------

In this work, we propose a method to create unconventional magnetic phases by using interfacial magnetic phase frustration as a driving mechanism. In the conventional geometrical frustration, as in spins in a triangular lattice,
couplings at various distances lead to unconventional phases due to competing tendencies. 
Here, in a multilayer geometry the competition arises at the center of a multilayer ensemble 
where dominant top and bottom states are not compatible with one another, thus stabilizing exotic central magnetic phases through interfacial phase frustration (IPF). 
In Fig.~\ref{schematic}, we show a sketch of our proposed platform, involving two $5d$ compounds at top and bottom, such as SrIrO$_3$ and CaIrO$_3$ with robust DMI coupling, as well as metallic lanthanum manganites La$_{1-x}$Sr$_{x}$MnO$_3$ (LSMO). The interface-induced DMI from the iridate generates, e.g., chiral magnetic phases in thin LSMO regions~\cite{Mohanta_PRB2019,Mohanta_CommPhys2020}. In this set up, the two LSMO layers can acquire a variety of magnetic phases including a spiral or a skyrmion crystal with various radii, depending on the doping level in the LSMO and the DMI strength at the iridate/LSMO interfaces. When two {\it different} magnetic phases are realized in the two LSMO layers, the interface between them will suffer the IPF that will produce an unconventional magnetic phase. We verified our proposal by performing Monte Carlo (MC) calculations using a spin Hamiltonian for a magnetic multilayer. We consider three cases and obtain three unconventional magnetic phases: (i) a checkerboard skyrmion crystal (CSkX) that appears from the competition between two orthogonally-aligned spin spirals, (ii) an incommensurate skyrmion stripe (ISkS) that arises from the competition between a spin spiral and a triangular skyrmion crystal, and (iii) a ferrimagnetic skyrmion crystal (FSkX) that arises from the competition between a triangular skyrmion crystal and a standard antiferromagnet. The emergent phases in the middle layer are unconventional, multi-\textit{q} magnetic textures and are predicted to stabilize in future experimental investigations. The CSkX is particularly interesting as it is created without any external magnetic field and is, therefore, promising for spintronic application. Furthermore, the CSkX produces a large topological Hall effect, induced by a nonzero scalar spin chirality, which is absent in the competing spin spirals. 
Thus, chirality arises from non-chiral components.

IPF paves the way to study artificial electrodynamics and unconventional magnetic excitations hosted by the frustration-stabilized magnetic phase in magnetic multilayers. It also broadens the class of chiral magnetic phases hosting skyrmions. This IPF can be realized generically using any two different competing magnetic phases. The primary advantage of the generation of unconventional magnetic orders via IPF is that the periodicity of the generated textures can be controlled by changing the constituent magnetic phases above and below. Additional potential material platforms for the experimental realization of the proposed IPF include superlattices of transition-metal oxide interfaces involving manganites in which a plethora of magnetic phases are already available for both Sr- and Ca-doped compounds~\cite{Yunoki_PRL2000,Moreo_PRL2000,Dagotto_PhyRep2001,Hotta_PRL2003}, chiral magnets {e.g.} MnSi and MnGe, as well as quasi-two-dimensional magnets, such as CrI$_3$ or Fe$_3$GeTe$_2$. Similar frustration-induced patterns can also be found at the domain walls between two or more magnetic phases~\cite{Schoenherr_NPhys2018}.

%---------------------------------------------
\begin{figure*}[t]
\begin{center}
\epsfig{file=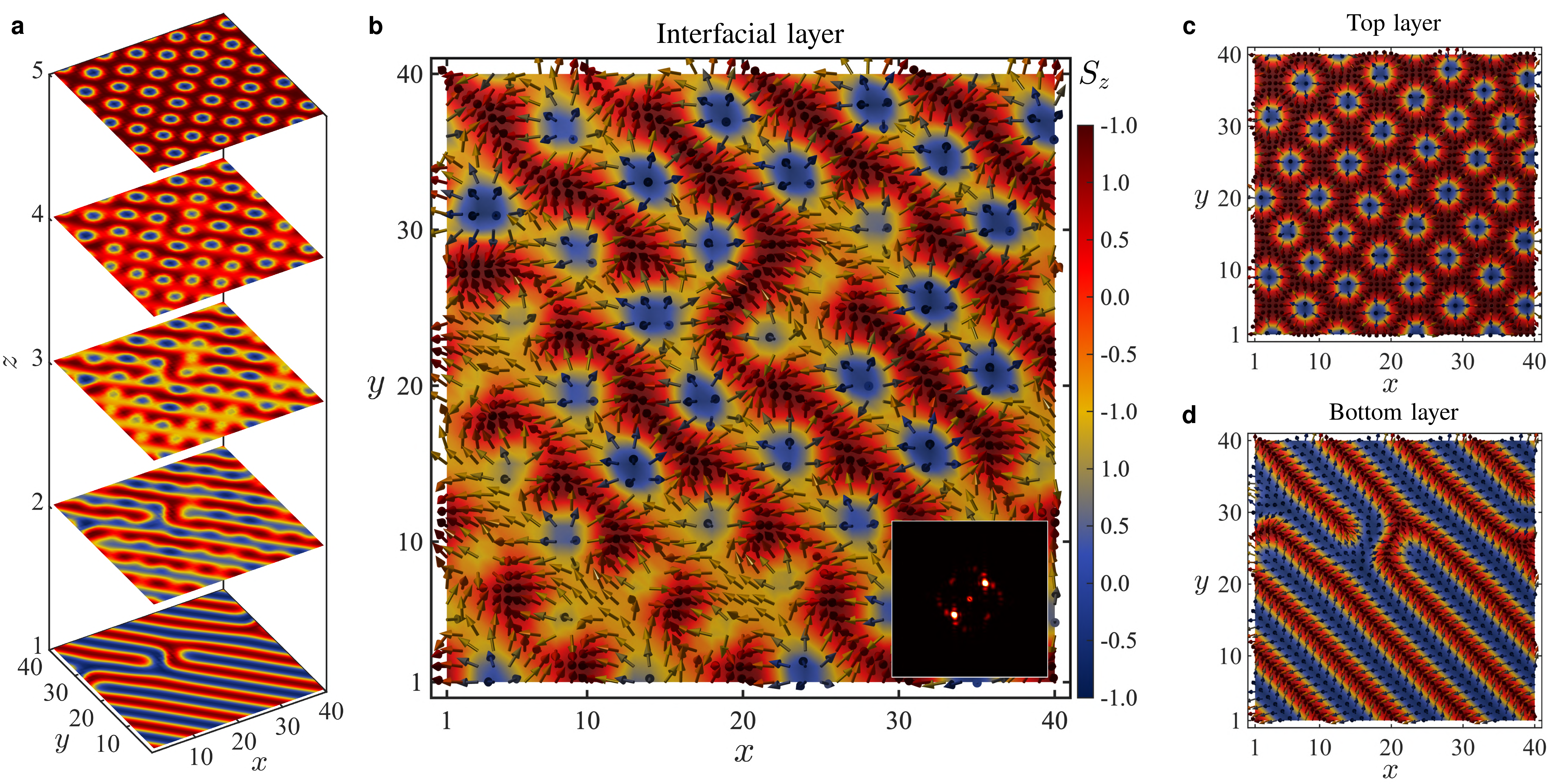,trim=0.0in 0.0in 0.0in 0.0in,clip=false, width=180mm}
\caption{{\bf Incommensurate skyrmion stripe.} {\bf a} Plot of the $z$-component of the spin $S_z$, obtained via MC annealing, in a five-layer heterostructure. {\bf b} Spin configuration of the IPF frustration-stabilized incommensurate skyrmion stripe (ISkS), obtained via MC annealing at the central layer ($z\!=\!3$). {\bf c} and {\bf d} are the spin configurations of a skyrmion crystal and a spin spiral at the top layer ($z\!=\!5$) and the bottom layer ($z\!=\!1$), respectively. The inset in {\bf b} shows the central-layer spin structure factor $S(\mathbf{q})$ within the first Brillouin zone, revealing a multi-\textit{q} magnetic order. The parameters used are $D\!=\!1.5J$, $A_z\!=\!0.3J$, $J\!=\!1$ (top layer), and $D\!=\!J$, $A_z\!=\!0$, $J\!=\!1$ (bottom layer).}
\label{isks}
\vspace{-2mm}
\end{center}
\end{figure*}
%---------------------------------------------

\noindent \small{\bf \\RESULTS}
\noindent {\bf \\Model\\}
Consider a multilayer heterostructure, with the top and the bottom layers hosting two different stable magnetic phases, such as the LSMO 1 and LSMO 2 layers in Fig.~\ref{schematic}. The magnetic phases of LSMO, above and below our focus -- the central layer, are influenced by the iridate/LSMO(1,2) interfaces. They can be described by the following Hamiltonian, with a ferromagnetic Heisenberg exchange coupling (because LSMO has a metallic ferromagnetic phase in a broad Sr-doping region), a DMI term from the iridates, and a single-ion uniaxial anisotropy 
\begin{align}
{\cal H}\!=&\!-J\! \sum_{\langle ij \rangle}\!\mathbf{S}_i \! \cdot \! \mathbf{S}_j \! 
-\!D\!\sum_{\langle ij \rangle} \!\hat{e}_{ij}\! \cdot \! \Big( \! \mathbf{S}_i \! \times \! \mathbf{S}_j \! \Big) 
-A_z\! \sum_{i} \! S_{zi}^2.
\end{align}
$\mathbf{S}_i$ represents a classical spin vector at site $i$, $J$ is the ferromagnetic exchange between neighboring spins, $D$ is the DMI coupling, $\hat{e}_{ij}$ is the unit vector connecting sites $i$ and $j$, and $A_z$ is the strength of uniaxial anisotropy along the $z$ direction.  The values of these parameters define separately the magnetic ground states in each layer. We keep the spin amplitude $S\!=\!1$ and exchange coupling amplitude $J\!=\!1$ throughout this paper, for simplicity. The individual layers are coupled via a ferromagnetic Heisenberg exchange coupling ${\cal H}\!=\!-J\! \sum_{\langle ij \rangle}\!\mathbf{S}_i \! \cdot \! \mathbf{S}_j $. We consider a five-layer heterostructure and perform MC annealing to obtain the spin configurations in all layers.

%---------------------------------------------
\begin{figure*}[t]
\begin{center}
\epsfig{file=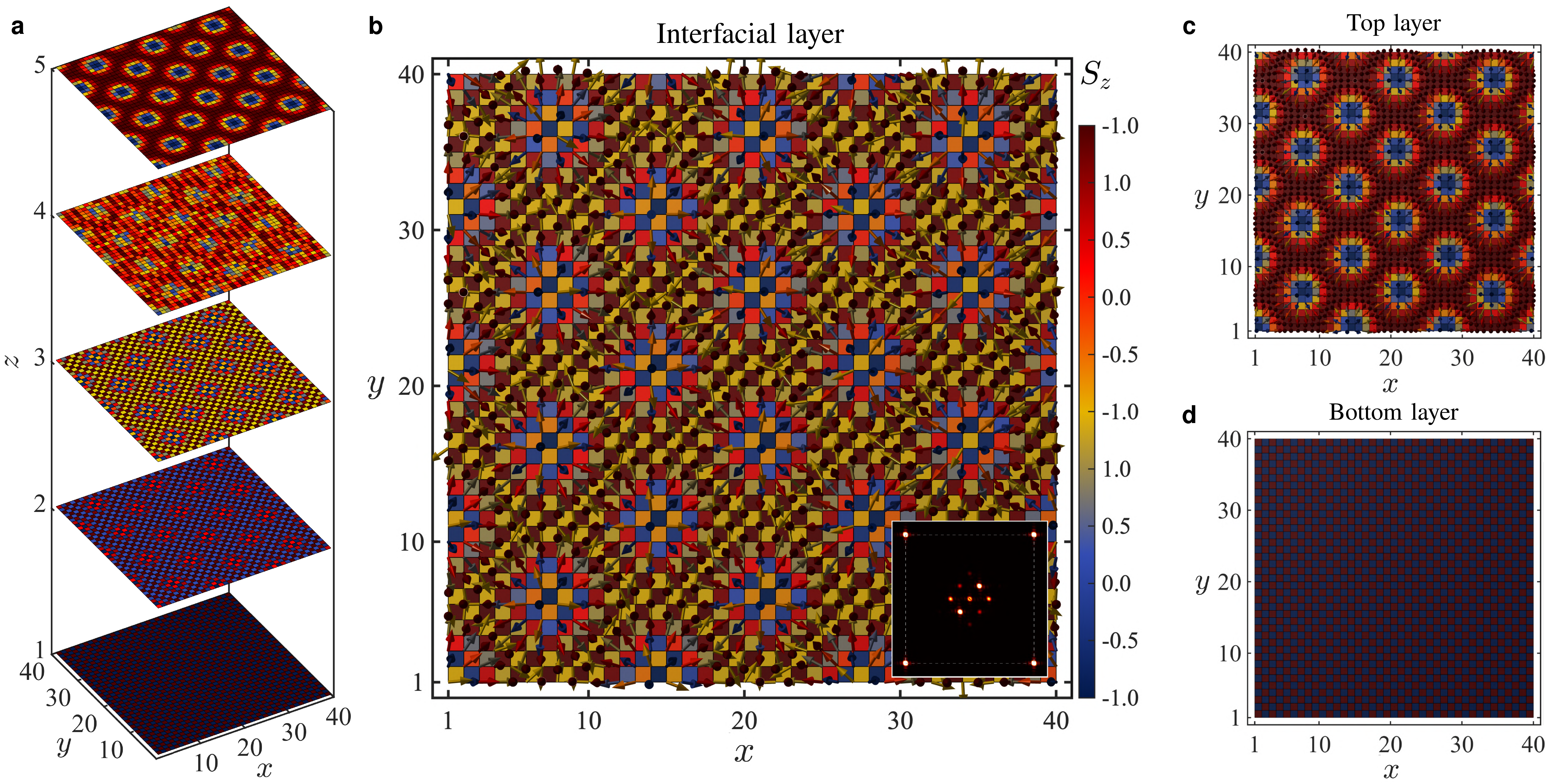,trim=0.0in 0.0in 0.0in 0.0in,clip=false, width=180mm}
\caption{{\bf Ferrimagnetic skyrmion crystal.} {\bf a} Plot of the $z$-component of the spin $S_z$, obtained via MC annealing, in a five-layer heterostructure. {\bf b} Spin configuration of the IPF frustration-stabilized ferrimagnetic skyrmion crystal (FSkX), obtained via MC annealing at the central layer ($z\!=\!3$). {\bf c} and {\bf d} are skyrmion crystal and antiferromagnet configurations at the top layer ($z\!=\!5$) and the bottom layer ($z\!=\!1$), respectively. The inset in {\bf b} shows the spin structure factor $S(\mathbf{q})$ within the first Brillouin zone (denoted by the dashed lines), revealing a multi-\textit{q} magnetic order. The parameters used are $D\!=\!J$, $A_z\!=\!0.24J$, $J\!=\!1$ (top layer), and $D\!=\!0$, $A_z\!=\!0$, $J\!=\!-1$ (bottom layer).}
\label{fskx}
\vspace{-4mm}
\end{center}
\end{figure*}
%---------------------------------------------

\noindent {\bf {\small \\Checkerboard skyrmion crystal.}}
We first consider two spin spirals as the top and bottom layers. The spin spiral phase, which appears with a finite DMI in the absence of any magnetic field or uniaxial anisotropy, has two degenerate spiral solutions, aligned mutually orthogonal to each other. One of these two preferred directions can be realized by directional tuning of a magnetic anisotropy~\cite{Garnier_JPM2020} or strain mismatching~\cite{Bishop_PRL2016}. In our MC calculations, these two spin spirals were spontaneously generated along the preferred directions by applying a small inplane anisotropy, described by ${\cal H}_{\rm in}^{\rm aniso}\!=\!-A_{\rm in}\! \sum_{i} \! (S_{xi}^2 \cos{\theta}+S_{yi}^2 \sin{\theta})$ along $\theta \!=\!45$$^{\circ}$ and $\theta \!=\!135$$^{\circ}$ at the top and the bottom layers, with $A_{\rm in} \!=\!0.05J$. Figures~\ref{cskx}{\bf a}-{\bf d} show the spin configurations at different layers of the considered five-layer heterostructure. The competition between the two spin spirals produces an unconventional checkerboard skyrmion crystal (CSkX). The spin structure factor $S(\mathbf{q})\!=\!(1/N)\sum_{ij}\langle \mathbf{S}_i \! \cdot \! \mathbf{S}_j  \rangle e^{-i\mathbf{q}\cdot \mathbf{r}_{ij}}$ -- where $\mathbf{r}_{ij}$ is the position vector connecting spins $\mathbf{S}_i$ and $\mathbf{S}_j$ and $N$ is the total number of lattice sites -- is shown in the inset of Fig.~\ref{cskx}{\bf b} and characterizes the double characteristic-momenta \textit{q} of the CSkX. Skyrmions are known to arise in non-centrosymmetric systems with a DMI and in centrosymmetric magnets with a geometrically-frustrated triangular lattice. It was unclear if skyrmions could appear without inversion symmetry breaking or geometrical frustration. Recent experiments reported a square  skyrmion crystal in GdRu$_2$Si$_2$, which is not explained by any of these two known mechanisms~\cite{Khanh_NatNano2020}. Our results may shed light on the origin of the reported square arrangement of skyrmions in centrosymmetric itinerant magnets.

\noindent {\bf {\small \\Incommensurate skyrmion stripe.}}
Next we focus on the second exotic state that we found, by considering a triangular skyrmion crystal at the top layer and a spin spiral at the bottom layer, as depicted in Figs.~\ref{isks}{\bf b} and {\bf c}, respectively. The lower critical magnetic field to obtain the skyrmion crystal depends on the DMI strength~\cite{Mohanta_PRB2020}, and therefore, in a heterostructure setting with two different values of the DMI strengths, the top and the bottom layers can host skyrmion crystal and spin spiral phases at a finite magnetic field. More interestingly, a skyrmion crystal can be stabilized by a uniaxial anisotropy without the need of any magnetic field, as observed, for example, in the  quasi-two-dimensional van der Waals magnet Fe$_3$GeTe$_2$~\cite{Yang_SciAdv2020}. In our MC calculation, the skyrmion crystal is realized in the presence of a perpendicular magnetic anisotropy at the top layer, at zero magnetic field. As shown in Fig.~\ref{fskx}, the spin configuration stabilized at the central layer by the IPF, due to the top skyrmion crystal and bottom spin spiral, is an unconventional multi-\textit{q} magnetic phase which we call incommensurate skyrmion stripe (ISkS) since the diagonal stripe order consists of four types of incommensurate skyrmionic patterns. Note the appearance of a skyrmionic diagonal pattern induced by the bottom layer which does not have skyrmions. Specifically, the spin structure factor $S(\mathbf{q})$, shown in the inset of Fig.~\ref{isks}{\bf b}, reveals multiple peaks. A careful inspection of this $S(\mathbf{q})$ profile indicates that among the nine peaks there are {\it five} characteristic momenta \textit{q} for this ISkS---the peak at zero momentum is for the ferromagnetic correlation, one of the two bright peaks inclined along 45$^{\circ}$ is for the diagonal stripe order, and three out of the six hexagon-shaped peaks are for the skyrmionic correlations.

\noindent {\bf {\small \\Ferrimagnetic skyrmion crystal.}}
In our third case, we consider a magnetic trilayer with a skyrmion crystal at the top layer and staggered antiferromagnetic order at the bottom layer. This situation can appear in our considered multilayer with different amount of doping in the top and bottom LSMO layers. As in the previous case, the skyrmion crystal is stabilized in our MC calculations by a finite DMI and an uniaxial anisotropy. On the other hand, the antiferromagnetic order is achieved in our spin Hamiltonian by a negative Heisenberg exchange constant $J\!=\!-1$, without DMI and uniaxial anisotropy. The magnetic configuration, generated at the central layer due to the IPF, is shown in Fig.~\ref{fskx}{\bf a}-{\bf b}, while the skyrmion crystal at the top layer and the antiferromagnetic configuration at the bottom layer are shown in Figs.~\ref{fskx}{\bf c} and {\bf d}. The IPF-stabilized texture contains remnants of the skyrmion crystal in the background of a ferrimagnetic order, thus the name ferrimagnetic skyrmion crystal (FSkX). The spin structure factor $S(\mathbf{q})$, shown in the inset of Fig.~\ref{fskx}{\bf b}, reveals a multi-\textit{q} pattern which has fingerprints of both the skyrmion crystal and the antiferromagnetic order.

\noindent {\bf {\small \\Topological Hall response.}}
Magnetic skyrmions are known to produce a topological Hall effect due to the emergent magnetic field $B_z\!=\!\frac{1}{2} \mathbf{n} \! \cdot \! \Big ( \! \partial_x \mathbf{n} \! \times \! \partial_y \mathbf{n} \Big)$ generated from the magnetic texture, where $\mathbf{n}$ represents the spin vector on the unit sphere created by the spin angles~\cite{Schulz_NatPhys2012,Nagaosa_NatNano2013}. Even though the proposed IPF mechanism is applicable to any type of magnetic phases---irrespective of its chiral nature --- because the three cases considered in this study are related to skyrmions -- either as top and bottom components or induced by IPF -- we analyzed the topological Hall response for the IPF-stabilized magnetic phases and compare with that of the constituent phases. 

%---------------------------------------------
\begin{figure}[t]
\begin{center}
\epsfig{file=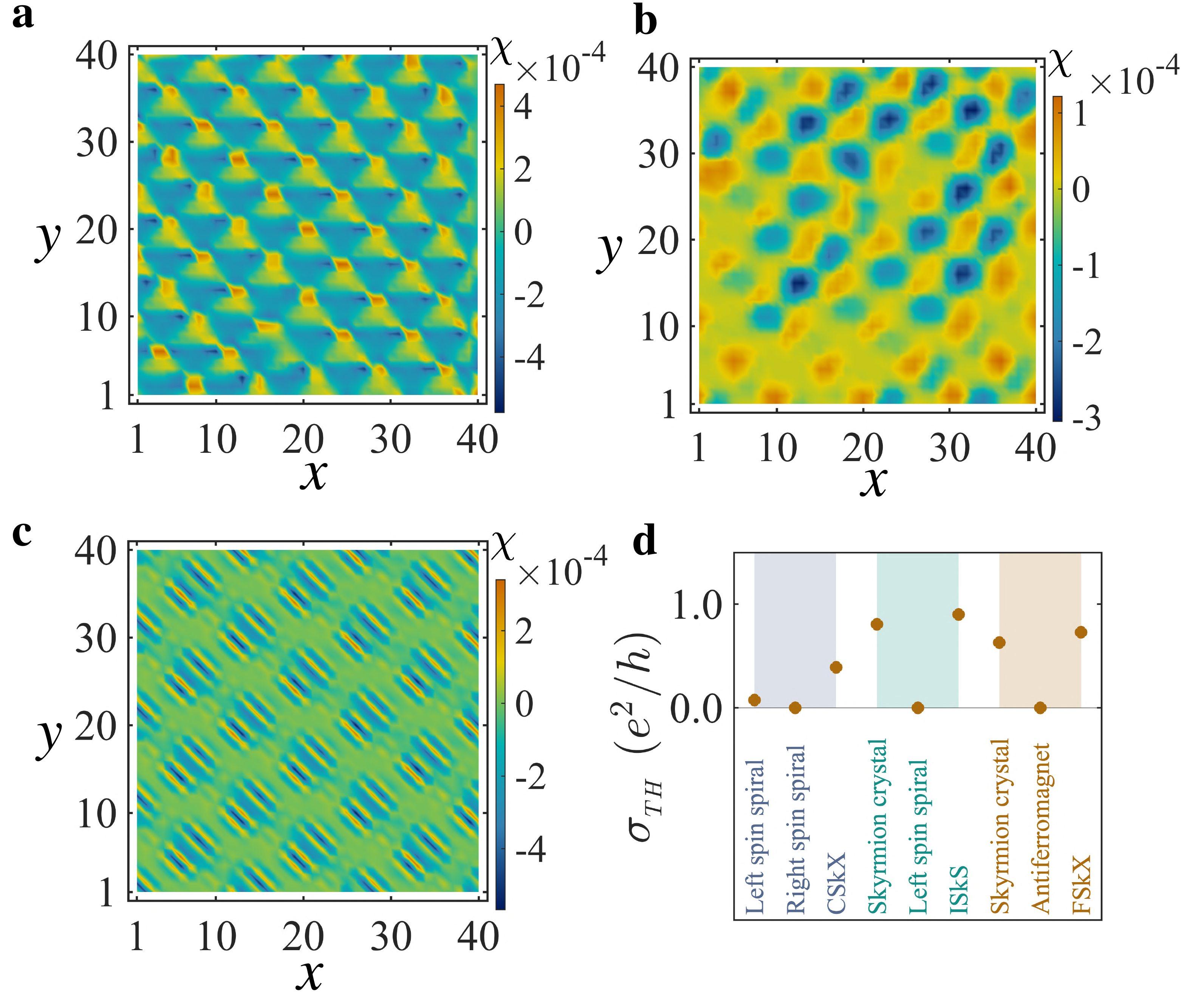,trim=0.0in 0.0in 0.0in 0.0in,clip=false, width=87mm}
\caption{{\bf Topological Hall effect.} Real-space profile of the scalar spin chirality $\chi (\mathbf{r}_i)$ for the three IPF-stabilized textures that appear in our MC simulations at the central layer of the considered multilayer: {\bf a} checkerboard skyrmion crystal (CSkX), {\bf b} incommensurate skyrmion stripe (ISkS), and {\bf c} ferrimagnetic skyrmion crystal (FSkX). {\bf d} Comparison of the topological Hall conductivity $\sigma_{_{\rm TH}}$ for these three unconventional magnetic phases and the constituent competing phases.}
\label{tHall}
\vspace{-2mm}
\end{center}
\end{figure}
%---------------------------------------------

To considering the impact that a chiral magnetic phase, such as those predicted, would have on the transverse Hall conductivity, it is useful to investigate the local scalar spin chirality $\chi (\mathbf{r}_i)$, which is a real-space equivalent of the Berry curvature that produces the Anomalous Hall Effect and is defined as
\begin{align}
\chi (\mathbf{r}_i)\!=\! \sum_{\langle ijk \rangle}\!\mathbf{S}_i \! \cdot \! (\mathbf{S}_j \! \times \! \mathbf{S}_k),
\end{align}
where $\langle ijk \rangle$ represents three neighboring lattice sites on a triangule of spins within the considered square lattice grid. The profile of $\chi (\mathbf{r}_i)$ for the three IPF-stabilized magnetic phases are depicted in Figs.~\ref{tHall}{\bf a}-{\bf c}, revealing a finite scalar spin chirality in all three phases. In order to compute the transverse Hall conductivity of these magnetic phases, we consider the following mobile fermions double-exchange Hamiltonian
\begin{align}
{\cal H}_{\rm DE}\!=\!-t\! \sum_{\langle ij \rangle}\! c_{i\sigma}^{\dagger}c_{j\sigma} -J_{\rm DE} \sum_{i,\sigma,\sigma^{\prime}} (\mathbf{S}_i \! \cdot \boldsymbol{\sigma}_{\sigma \sigma^{\prime}})c_{i\sigma}^{\dagger}c_{i\sigma^{\prime}}
\end{align}
where $t$ is the nearest-neighbor electron hopping energy and $J_{\rm DE}$ is the Hund exchange coupling strength between the itinerant electron spin $\boldsymbol{\sigma}$ and the localized spin $\mathbf{S}_i$ at a particular layer. In other words, we consider mobile electrons in the background of the fixed IPF textures. The topological Hall conductivity is obtained via the canonical Kubo formula
\begin{align}
\sigma_{_{\rm TH}} \!=\! \frac{e^2}{h}\frac{2\pi}{N} \! \sum_{\epsilon_m \neq \epsilon_n} \! \frac{f_m-f_n}{(\epsilon_m \!-\! \epsilon_n)^2 \!+\! \eta^2}{\rm Im} \Big( \langle m | \hat{j}_x | n \rangle \langle n | \hat{j}_y | m \rangle \Big),
\end{align}
where $f_m$ is the Fermi function at temperature $T$ and energy eigenvalue $\epsilon_m$, $\hat{j}_x$ is the current operator along the $x$ direction, $|m\rangle$ is the $m^{\rm th}$ eigenvector of ${\cal H}_{\rm DE}$, and $\eta$ is the relaxation rate. We used $t\!=\!1$, $J_{\rm DE}\!=\!1$, and $\eta \!=\!0.1$ for the present analysis, with no qualitative difference in the description for other choices, as discussed before~\cite{Mohanta_PRB2019}. In Fig.~\ref{tHall}{\bf d}, we present the topological Hall conductivity for the magnetic textures that we consider in the three cases discussed above. Clearly, $\sigma_{_{\rm TH}}$ is finite for all three chiral magnetic phases stabilized by the IPF. Notably, $\sigma_{_{\rm TH}}$ acquires a finite value for the CSkX phase, even though the contribution from the spin spirals on the top and the bottom layers are nearly zero. This finding, therefore, could be "fingerprints" useful in deciphering topological Hall signals from magnetic multilayers where the top and bottom spin spirals are oriented along different directions. We also note that $\sigma_{_{\rm TH}}$ from the other two IPF-stabilized magnetic phases, \textit{viz.}, ISkS and FSkX, are comparable to that from the constituent skyrmion crystal, which implies that the resultant topological Hall effect will be nearly doubled than that of the expected value from only the skyrmion crystal. The appearance of topological Hall effect from non-skyrmionic magnetic phases is an interesting topic of recent discussion~\cite{Mohanta_PRB2020,Nakane_PRB2020}. The current findings support these theoretical predictions, albeit via a totally different IPF platform, and seek further experimental studies in this direction---exploring topological Hall effect from non-skyrmionic chiral magnetic textures. 

In our five-layer structures, the extreme top and the bottom layers represent the stable constituent magnetic phases, while the central layer reveals the IPF-induced magnetic phase. The second and fourth layers show deformed intermediate magnetic configurations, as shown in Figs.~\ref{cskx}{\bf a}, ~\ref{isks}{\bf a}, and ~\ref{fskx}{\bf a}, and these layers can also possess a finite scalar spin chirality or topological charge. We also noted that despite the little imperfections in the top and bottom spin configurations that we obtained via MC annealing, the IPF-induced pattern at the central layer defines an unconventional state.

\noindent \small{\bf {\\DISCUSSION\\}}
The interfacial DMI, that is primarily responsible for the chiral magnetic interactions, can be controlled in $3d$--$5d$ transition-metal-oxide-based superlattices, such as those made of manganite/iridate interfaces by engineering the interfacial inversion symmetry~\cite{Yi_PNAS2016,Skoropata_SciAdv2020}. With advancements in the deposition techniques, the layer thickness can also tune the Rashba-type interfacial DMI, Dresselhaus-type bulk DMI, and the anisotropy in the B20 structures, such as CrGe/MnGe/FeGe and their superlattices~\cite{Kawakami_JCG2017}. Additional strain control in these B20 structures can, furthermore, create complex spin textures from the superposition of different helical domains~\cite{Repicky_Science2021}. Therefore, the CrGe/MnGe/FeGe compounds could be used for the experimental realization of the discussed IPF and generate unconventional magnetic phases. Besides transition metal oxide interfaces and chiral magnets hosting skyrmions, the IPF may be explored in quasi-two-dimensional magnets, such as CrI$_3$, or Fe-based van der Waals magnets, such as Fe$_3$GeTe$_2$.

The appearance of a multi-\textit{q} order is a generic property of the IPF-induced phases. These multi-\textit{q} phases can harbor soft magnetic modes that can be controlled by an electric field, leading to a characteristic magnetoelectric effect~\cite{Leonov_NatComm2015}. Furthermore, due to the frustration, these multi-\textit{q} phases may allow skyrmions and anti-skyrmions to be present together in an exotic thermodynamic phase that can host magnetic monopoles~\cite{Okubo_PRL2012}.

The IPF-induced phases are, in general, expected to appear below the critical temperatures of the constituent magnetic phases in the multilayer, whichever is lesser. The chiral magnetic phases including the SkX commonly appear below 100 K, as revealed by topological Hall effect measurements~\cite{PhysRevLett.102.186602,Matsuno_sciadv,Wang2018,Vistoli2019}. Therefore, the three unconventional SkX phases that we discuss here are also expected to appear within a similar temperature range. Nonetheless, by choosing constituent phases that persist at higher temperatures, it is possible to realize unconventional magnetic phases at higher temperatures.

There is a growing interest in three-dimensional chiral magnetic phases for the urge to understand the complexity arising from topology and frustration in three dimensions, and also for their potential spintronic applications~\cite{Pacheco_NatComm2017, Gao_Nature2020, Kent_NatCommun2021}. With the layer control that is predicted here, it would be possible to achieve unconventional magnetic phases with potential functionalities. The synthetic spin-orbit coupling available in these exotic skyrmion crystals may also be useful in fault-tolerant topological quantum computing~\cite{Mohanta_CommPhys2021, Mascot_NPJQM2021}. 

To summarize, we proposed a method to stabilize never-observed-before magnetic phases by using interfacial phase frustration in multilayers. As example, we obtained three exotic magnetic phases in our MC calculations, generated by this interfacial phase frustration. We hope that the proposed frustration mechanism will open a path, aiding the search of unconventional magnetism in three dimensions.

\noindent \small{\bf {\\METHODS\\}}
\noindent \small{\bf {Monte Carlo simulations\\}}
We perform Monte Carlo (MC) annealing to obtain the ground state spin configuration in the discussed magnetic multilayer. We consider periodic boundary conditions in plane and use the Metropolis spin update procedure. The ground state magnetic configurations at the top two and the bottom two layers in our five-layer structure were first stabilized by performing MC annealing, without any communication between the top two and bottom two layers. Next, we switch on the interlayer coupling between all the layers and perform MC annealing in the entire multilayer. During this process, we start from a completely random spin configuration at the middle layer. In the MC annealing processes, we start at a high temperature $T\!=\!5J$ and very slowly reduce the temperature down to $T\!=\!0.001J$. At each temperature, a large number of MC spin updates, typically of the order of $10^{11}$, were performed to avoid being trapped in metastable states.

\noindent {\bf {\small \\Data availability\\}} Data are available from the corresponding author upon reasonable request.

\noindent {\bf {\small \\Code availability\\}} Simulation codes are available from the corresponding author upon reasonable
request.

\def\bibsection{\section*{\refname}} 
\bibliographystyle{naturemag}
%\bibliography{Ref}

\begin{thebibliography}{10}
\expandafter\ifx\csname url\endcsname\relax
  \def\url#1{\texttt{#1}}\fi
\expandafter\ifx\csname urlprefix\endcsname\relax\def\urlprefix{URL: }\fi
\providecommand{\bibinfo}[2]{#2}
\providecommand{\eprint}[2][]{\url{#2}}

\bibitem{Bramwell_Science2001}
\bibinfo{author}{Bramwell, S.~T.} \& \bibinfo{author}{Gingras, M. J.~P.}
\newblock \bibinfo{title}{Spin ice state in frustrated magnetic pyrochlore
  materials}.
\newblock \textit{\bibinfo{journal}{Science}} \textbf{\bibinfo{volume}{294}},
  \bibinfo{pages}{1495--1501} (\bibinfo{year}{2001}).
\newblock
  \urlprefix\url{https://www.science.org/doi/abs/10.1126/science.1064761}.

\bibitem{Balents2010}
\bibinfo{author}{Balents, L.}
\newblock \bibinfo{title}{Spin liquids in frustrated magnets}.
\newblock \textit{\bibinfo{journal}{Nature}} \textbf{\bibinfo{volume}{464}},
  \bibinfo{pages}{199–208} (\bibinfo{year}{2010}).
\newblock \urlprefix\url{https://doi.org/10.1038/nature08917}.

\bibitem{Kitaev2003}
\bibinfo{author}{Kitaev, A.}
\newblock \bibinfo{title}{Fault-tolerant quantum computation by anyons}.
\newblock \textit{\bibinfo{journal}{Ann. Phys.}}
  \textbf{\bibinfo{volume}{303}}, \bibinfo{pages}{2--30}
  (\bibinfo{year}{2003}).
\newblock
  \urlprefix\url{https://www.sciencedirect.com/science/article/pii/S0003491602000180}.

\bibitem{Savary_RPP2016}
\bibinfo{author}{Savary, L.} \& \bibinfo{author}{Balents, L.}
\newblock \bibinfo{title}{Quantum spin liquids: {A} review}.
\newblock \textit{\bibinfo{journal}{Rep. Prog. Phys.}}
  \textbf{\bibinfo{volume}{80}}, \bibinfo{pages}{016502}
  (\bibinfo{year}{2016}).
\newblock \urlprefix\url{https://doi.org/10.1088/0034-4885/80/1/016502}.

\bibitem{Batista_RPP2016}
\bibinfo{author}{Batista, C.~D.}, \bibinfo{author}{Lin, S.-Z.},
  \bibinfo{author}{Hayami, S.} \& \bibinfo{author}{Kamiya, Y.}
\newblock \bibinfo{title}{Frustration and chiral orderings in correlated
  electron systems}.
\newblock \textit{\bibinfo{journal}{Rep. Prog. Phys.}}
  \textbf{\bibinfo{volume}{79}}, \bibinfo{pages}{084504}
  (\bibinfo{year}{2016}).
\newblock \urlprefix\url{https://doi.org/10.1088/0034-4885/79/8/084504}.

\bibitem{Taguchi_Science2001}
\bibinfo{author}{Taguchi, Y.}, \bibinfo{author}{Oohara, Y.},
  \bibinfo{author}{Yoshizawa, H.}, \bibinfo{author}{Nagaosa, N.} \&
  \bibinfo{author}{Tokura, Y.}
\newblock \bibinfo{title}{{Spin chirality, Berry phase, and anomalous Hall
  effect in a frustrated ferromagnet}}.
\newblock \textit{\bibinfo{journal}{Science}} \textbf{\bibinfo{volume}{291}},
  \bibinfo{pages}{2573--2576} (\bibinfo{year}{2001}).
\newblock
  \urlprefix\url{https://www.science.org/doi/abs/10.1126/science.1058161}.

\bibitem{Grohol_NatMat2005}
\bibinfo{author}{Grohol, D.} \textit{et~al.}
\newblock \bibinfo{title}{Spin chirality on a two-dimensional frustrated
  lattice}.
\newblock \textit{\bibinfo{journal}{Nat. Mater.}} \textbf{\bibinfo{volume}{4}},
  \bibinfo{pages}{323–328} (\bibinfo{year}{2005}).
\newblock \urlprefix\url{https://doi.org/10.1038/nmat1353}.

\bibitem{Schulz_NatPhys2012}
\bibinfo{author}{Schulz, T.} \textit{et~al.}
\newblock \bibinfo{title}{Emergent electrodynamics of skyrmions in a chiral
  magnet}.
\newblock \textit{\bibinfo{journal}{Nat. Phys.}} \textbf{\bibinfo{volume}{8}},
  \bibinfo{pages}{301–304} (\bibinfo{year}{2012}).
\newblock \urlprefix\url{https://doi.org/10.1038/nphys2231}.

\bibitem{Nagaosa_NatNano2013}
\bibinfo{author}{Nagaosa, N.} \& \bibinfo{author}{Tokura, Y.}
\newblock \bibinfo{title}{Topological properties and dynamics of magnetic
  skyrmions}.
\newblock \textit{\bibinfo{journal}{Nat. Nanotechnol.}}
  \textbf{\bibinfo{volume}{8}}, \bibinfo{pages}{899–911}
  (\bibinfo{year}{2013}).
\newblock \urlprefix\url{https://doi.org/10.1038/nnano.2013.243}.

\bibitem{Fert_NarRevMat2017}
\bibinfo{author}{Fert, A.}, \bibinfo{author}{Reyren, N.} \&
  \bibinfo{author}{Cros, V.}
\newblock \bibinfo{title}{Magnetic skyrmions: {A}dvances in physics and
  potential applications}.
\newblock \textit{\bibinfo{journal}{Nat. Rev. Mater.}}
  \textbf{\bibinfo{volume}{2}}, \bibinfo{pages}{17031} (\bibinfo{year}{2017}).
\newblock \urlprefix\url{https://doi.org/10.1038/natrevmats.2017.31}.

\bibitem{Rossler_Nature2006}
\bibinfo{author}{R{\"o}{\ss}ler, U.~K.}, \bibinfo{author}{Bogdanov, A.~N.} \&
  \bibinfo{author}{Pfleiderer, C.}
\newblock \bibinfo{title}{Spontaneous skyrmion ground states in magnetic
  metals}.
\newblock \textit{\bibinfo{journal}{Nature}} \textbf{\bibinfo{volume}{442}},
  \bibinfo{pages}{797–801} (\bibinfo{year}{2006}).
\newblock \urlprefix\url{https://doi.org/10.1038/nature05056}.

\bibitem{Muhlbauer_Science2009}
\bibinfo{author}{Mühlbauer, S.} \textit{et~al.}
\newblock \bibinfo{title}{Skyrmion lattice in a chiral magnet}.
\newblock \textit{\bibinfo{journal}{Science}} \textbf{\bibinfo{volume}{323}},
  \bibinfo{pages}{915--919} (\bibinfo{year}{2009}).
\newblock
  \urlprefix\url{https://www.science.org/doi/abs/10.1126/science.1166767}.

\bibitem{Yu_Nature2010}
\bibinfo{author}{Yu, X.~Z.} \textit{et~al.}
\newblock \bibinfo{title}{Real-space observation of a two-dimensional skyrmion
  crystal}.
\newblock \textit{\bibinfo{journal}{Nature}} \textbf{\bibinfo{volume}{465}},
  \bibinfo{pages}{901–904} (\bibinfo{year}{2010}).
\newblock \urlprefix\url{https://doi.org/10.1038/nature09124}.

\bibitem{Balents_PRL2014}
\bibinfo{author}{Li, X.}, \bibinfo{author}{Liu, W.~V.} \&
  \bibinfo{author}{Balents, L.}
\newblock \bibinfo{title}{Spirals and skyrmions in two dimensional oxide
  heterostructures}.
\newblock \textit{\bibinfo{journal}{Phys. Rev. Lett.}}
  \textbf{\bibinfo{volume}{112}}, \bibinfo{pages}{067202}
  (\bibinfo{year}{2014}).
\newblock
  \urlprefix\url{https://link.aps.org/doi/10.1103/PhysRevLett.112.067202}.

\bibitem{Das_Nature2019}
\bibinfo{author}{Das, S.} \textit{et~al.}
\newblock \bibinfo{title}{Observation of room-temperature polar skyrmions}.
\newblock \textit{\bibinfo{journal}{Nature}} \textbf{\bibinfo{volume}{568}},
  \bibinfo{pages}{368–372} (\bibinfo{year}{2019}).
\newblock \urlprefix\url{https://doi.org/10.1038/s41586-019-1092-8}.

\bibitem{Mohanta_PRB2019}
\bibinfo{author}{Mohanta, N.}, \bibinfo{author}{Dagotto, E.} \&
  \bibinfo{author}{Okamoto, S.}
\newblock \bibinfo{title}{{Topological Hall effect and emergent skyrmion
  crystal at manganite-iridate oxide interfaces}}.
\newblock \textit{\bibinfo{journal}{Phys. Rev. B}}
  \textbf{\bibinfo{volume}{100}}, \bibinfo{pages}{064429}
  (\bibinfo{year}{2019}).
\newblock \urlprefix\url{https://link.aps.org/doi/10.1103/PhysRevB.100.064429}.

\bibitem{Okubo_PRL2012}
\bibinfo{author}{Okubo, T.}, \bibinfo{author}{Chung, S.} \&
  \bibinfo{author}{Kawamura, H.}
\newblock \bibinfo{title}{Multiple-$q$ states and the skyrmion lattice of the
  triangular-lattice {H}eisenberg antiferromagnet under magnetic fields}.
\newblock \textit{\bibinfo{journal}{Phys. Rev. Lett.}}
  \textbf{\bibinfo{volume}{108}}, \bibinfo{pages}{017206}
  (\bibinfo{year}{2012}).
\newblock
  \urlprefix\url{https://link.aps.org/doi/10.1103/PhysRevLett.108.017206}.

\bibitem{Leonov_NatComm2015}
\bibinfo{author}{Leonov, A.~O.} \& \bibinfo{author}{Mostovoy, M.}
\newblock \bibinfo{title}{Multiply periodic states and isolated skyrmions in an
  anisotropic frustrated magnet}.
\newblock \textit{\bibinfo{journal}{Nat. Commun.}}
  \textbf{\bibinfo{volume}{6}}, \bibinfo{pages}{8275} (\bibinfo{year}{2015}).
\newblock \urlprefix\url{https://doi.org/10.1038/ncomms9275}.

\bibitem{Ozawa_PRL2017}
\bibinfo{author}{Ozawa, R.}, \bibinfo{author}{Hayami, S.} \&
  \bibinfo{author}{Motome, Y.}
\newblock \bibinfo{title}{Zero-field skyrmions with a high topological number
  in itinerant magnets}.
\newblock \textit{\bibinfo{journal}{Phys. Rev. Lett.}}
  \textbf{\bibinfo{volume}{118}}, \bibinfo{pages}{147205}
  (\bibinfo{year}{2017}).
\newblock
  \urlprefix\url{https://link.aps.org/doi/10.1103/PhysRevLett.118.147205}.

\bibitem{Wang_PRL2020}
\bibinfo{author}{Wang, Z.}, \bibinfo{author}{Su, Y.}, \bibinfo{author}{Lin,
  S.-Z.} \& \bibinfo{author}{Batista, C.~D.}
\newblock \bibinfo{title}{Skyrmion crystal from {RKKY} interaction mediated by
  {2D} electron gas}.
\newblock \textit{\bibinfo{journal}{Phys. Rev. Lett.}}
  \textbf{\bibinfo{volume}{124}}, \bibinfo{pages}{207201}
  (\bibinfo{year}{2020}).
\newblock
  \urlprefix\url{https://link.aps.org/doi/10.1103/PhysRevLett.124.207201}.

\bibitem{Park_NatNano2014}
\bibinfo{author}{Park, H.~S.} \textit{et~al.}
\newblock \bibinfo{title}{Observation of the magnetic flux and
  three-dimensional structure of skyrmion lattices by electron holography}.
\newblock \textit{\bibinfo{journal}{Nat. Nanotechnol.}}
  \textbf{\bibinfo{volume}{9}}, \bibinfo{pages}{337–342}
  (\bibinfo{year}{2014}).
\newblock \urlprefix\url{https://doi.org/10.1038/nnano.2014.52}.

\bibitem{Zheng2018}
\bibinfo{author}{Zheng, F.} \textit{et~al.}
\newblock \bibinfo{title}{Experimental observation of chiral magnetic bobbers
  in {B20-type FeGe}}.
\newblock \textit{\bibinfo{journal}{Nat. Nanotechnol.}}
  \textbf{\bibinfo{volume}{13}}, \bibinfo{pages}{451–455}
  (\bibinfo{year}{2018}).
\newblock \urlprefix\url{https://doi.org/10.1038/s41565-018-0093-3}.

\bibitem{Kent_NatCommun2021}
\bibinfo{author}{Kent, N.} \textit{et~al.}
\newblock \bibinfo{title}{Creation and observation of {H}opfions in magnetic
  multilayer systems}.
\newblock \textit{\bibinfo{journal}{Nat. Commun.}}
  \textbf{\bibinfo{volume}{12}}, \bibinfo{pages}{1562} (\bibinfo{year}{2021}).
\newblock \urlprefix\url{https://doi.org/10.1038/s41467-021-21846-5}.

\bibitem{Mohanta_CommPhys2020}
\bibinfo{author}{Mohanta, N.}, \bibinfo{author}{Christianson, A.~D.},
  \bibinfo{author}{Okamoto, S.} \& \bibinfo{author}{Dagotto, E.}
\newblock \bibinfo{title}{Signatures of a liquid-crystal transition in
  spin-wave excitations of skyrmions}.
\newblock \textit{\bibinfo{journal}{Commun. Phys.}}
  \textbf{\bibinfo{volume}{3}}, \bibinfo{pages}{229} (\bibinfo{year}{2020}).
\newblock \urlprefix\url{https://doi.org/10.1038/s42005-020-00489-w}.

\bibitem{Yunoki_PRL2000}
\bibinfo{author}{Yunoki, S.}, \bibinfo{author}{Hotta, T.} \&
  \bibinfo{author}{Dagotto, E.}
\newblock \bibinfo{title}{Ferromagnetic, $\mathit{A}$-type, and charge-ordered
  $\mathrm{CE}$-type states in doped manganites using {J}ahn-{T}eller phonons}.
\newblock \textit{\bibinfo{journal}{Phys. Rev. Lett.}}
  \textbf{\bibinfo{volume}{84}}, \bibinfo{pages}{3714--3717}
  (\bibinfo{year}{2000}).
\newblock \urlprefix\url{https://link.aps.org/doi/10.1103/PhysRevLett.84.3714}.

\bibitem{Moreo_PRL2000}
\bibinfo{author}{Moreo, A.}, \bibinfo{author}{Mayr, M.},
  \bibinfo{author}{Feiguin, A.}, \bibinfo{author}{Yunoki, S.} \&
  \bibinfo{author}{Dagotto, E.}
\newblock \bibinfo{title}{Giant cluster coexistence in doped manganites and
  other compounds}.
\newblock \textit{\bibinfo{journal}{Phys. Rev. Lett.}}
  \textbf{\bibinfo{volume}{84}}, \bibinfo{pages}{5568--5571}
  (\bibinfo{year}{2000}).
\newblock \urlprefix\url{https://link.aps.org/doi/10.1103/PhysRevLett.84.5568}.

\bibitem{Dagotto_PhyRep2001}
\bibinfo{author}{Dagotto, E.}, \bibinfo{author}{Hotta, T.} \&
  \bibinfo{author}{Moreo, A.}
\newblock \bibinfo{title}{Colossal magnetoresistant materials: the key role of
  phase separation}.
\newblock \textit{\bibinfo{journal}{Phys. Rep.}}
  \textbf{\bibinfo{volume}{344}}, \bibinfo{pages}{1--153}
  (\bibinfo{year}{2001}).
\newblock
  \urlprefix\url{https://www.sciencedirect.com/science/article/pii/S0370157300001216}.

\bibitem{Hotta_PRL2003}
\bibinfo{author}{Hotta, T.}, \bibinfo{author}{Moraghebi, M.},
  \bibinfo{author}{Feiguin, A.}, \bibinfo{author}{Moreo, A.},
  \bibinfo{author}{Yunoki, S.} \& \bibinfo{author}{Dagotto, E.}
\newblock \bibinfo{title}{Unveiling new magnetic phases of undoped and doped
  manganites}.
\newblock \textit{\bibinfo{journal}{Phys. Rev. Lett.}}
  \textbf{\bibinfo{volume}{90}}, \bibinfo{pages}{247203}
  (\bibinfo{year}{2003}).
\newblock
  \urlprefix\url{https://link.aps.org/doi/10.1103/PhysRevLett.90.247203}.

\bibitem{Schoenherr_NPhys2018}
\bibinfo{author}{Schoenherr, P.} \textit{et~al.}
\newblock \bibinfo{title}{Topological domain walls in helimagnets}.
\newblock \textit{\bibinfo{journal}{Nat. Phys.}} \textbf{\bibinfo{volume}{14}},
  \bibinfo{pages}{465–468} (\bibinfo{year}{2018}).
\newblock \urlprefix\url{https://doi.org/10.1038/s41567-018-0056-5}.

\bibitem{Garnier_JPM2020}
\bibinfo{author}{Garnier, L.-C.} \textit{et~al.}
\newblock \bibinfo{title}{{Stripe domains reorientation in ferromagnetic films
  with perpendicular magnetic anisotropy}}.
\newblock \textit{\bibinfo{journal}{J. Phys.: Mater.}}
  \textbf{\bibinfo{volume}{3}}, \bibinfo{pages}{024001} (\bibinfo{year}{2020}).
\newblock \urlprefix\url{https://doi.org/10.1088/2515-7639/ab6ea5}.

\bibitem{Bishop_PRL2016}
\bibinfo{author}{Bishop, C.~B.}, \bibinfo{author}{Moreo, A.} \&
  \bibinfo{author}{Dagotto, E.}
\newblock \bibinfo{title}{Bicollinear antiferromagnetic order, monoclinic
  distortion, and reversed resistivity anisotropy in {FeTe} as a result of
  spin-lattice coupling}.
\newblock \textit{\bibinfo{journal}{Phys. Rev. Lett.}}
  \textbf{\bibinfo{volume}{117}}, \bibinfo{pages}{117201}
  (\bibinfo{year}{2016}).
\newblock
  \urlprefix\url{https://link.aps.org/doi/10.1103/PhysRevLett.117.117201}.

\bibitem{Khanh_NatNano2020}
\bibinfo{author}{Khanh, N.~D.} \textit{et~al.}
\newblock \bibinfo{title}{Nanometric square skyrmion lattice in a
  centrosymmetric tetragonal magnet}.
\newblock \textit{\bibinfo{journal}{Nat. Nanotechnol.}}
  \textbf{\bibinfo{volume}{15}}, \bibinfo{pages}{444–449}
  (\bibinfo{year}{2020}).
\newblock \urlprefix\url{https://doi.org/10.1038/s41565-020-0684-7}.

\bibitem{Mohanta_PRB2020}
\bibinfo{author}{Mohanta, N.}, \bibinfo{author}{Okamoto, S.} \&
  \bibinfo{author}{Dagotto, E.}
\newblock \bibinfo{title}{Planar topological {H}all effect from conical spin
  spirals}.
\newblock \textit{\bibinfo{journal}{Phys. Rev. B}}
  \textbf{\bibinfo{volume}{102}}, \bibinfo{pages}{064430}
  (\bibinfo{year}{2020}).
\newblock \urlprefix\url{https://link.aps.org/doi/10.1103/PhysRevB.102.064430}.

\bibitem{Yang_SciAdv2020}
\bibinfo{author}{Yang, M.} \textit{et~al.}
\newblock \bibinfo{title}{Creation of skyrmions in van der {W}aals ferromagnet
  {Fe}$_3${GeTe}$_2$ on ({Co}/{Pd})$_n$ superlattice}.
\newblock \textit{\bibinfo{journal}{Sci. Adv.}} \textbf{\bibinfo{volume}{6}},
  \bibinfo{pages}{eabb5157} (\bibinfo{year}{2020}).
\newblock
  \urlprefix\url{https://www.science.org/doi/abs/10.1126/sciadv.abb5157}.

\bibitem{Nakane_PRB2020}
\bibinfo{author}{Nakane, J.~J.}, \bibinfo{author}{Nakazawa, K.} \&
  \bibinfo{author}{Kohno, H.}
\newblock \bibinfo{title}{Topological {H}all effect in weakly canted
  antiferromagnets}.
\newblock \textit{\bibinfo{journal}{Phys. Rev. B}}
  \textbf{\bibinfo{volume}{101}}, \bibinfo{pages}{174432}
  (\bibinfo{year}{2020}).
\newblock \urlprefix\url{https://link.aps.org/doi/10.1103/PhysRevB.101.174432}.

\bibitem{Yi_PNAS2016}
\bibinfo{author}{Yi, D.} \textit{et~al.}
\newblock \bibinfo{title}{Atomic-scale control of magnetic anisotropy via novel
  spin-orbit coupling effect in {L}a$_{2/3}${S}r$_{1/3}${M}no$_3$/{S}r{I}r{O}3
  superlattices}.
\newblock \textit{\bibinfo{journal}{Proc. Natl. Acad. Sci.}}
  \textbf{\bibinfo{volume}{113}}, \bibinfo{pages}{6397--6402}
  (\bibinfo{year}{2016}).
\newblock \urlprefix\url{https://www.pnas.org/content/113/23/6397}.

\bibitem{Skoropata_SciAdv2020}
\bibinfo{author}{Skoropata, E.} \textit{et~al.}
\newblock \bibinfo{title}{Interfacial tuning of chiral magnetic interactions
  for large topological {H}all effects in {L}a{M}n{O}$_3$/{S}r{I}r{O}$_3$
  heterostructures}.
\newblock \textit{\bibinfo{journal}{Sci. Adv.}} \textbf{\bibinfo{volume}{6}},
  \bibinfo{pages}{eaaz3902} (\bibinfo{year}{2020}).
\newblock
  \urlprefix\url{https://www.science.org/doi/abs/10.1126/sciadv.aaz3902}.

\bibitem{Kawakami_JCG2017}
\bibinfo{author}{Ahmed, A.~S.}, \bibinfo{author}{Esser, B.~D.},
  \bibinfo{author}{Rowland, J.}, \bibinfo{author}{McComb, D.~W.} \&
  \bibinfo{author}{Kawakami, R.~K.}
\newblock \bibinfo{title}{Molecular beam epitaxy growth of [{CrGe/MnGe/FeGe}]
  superlattices: {T}oward artificial {B20} skyrmion materials with tunable
  interactions}.
\newblock \textit{\bibinfo{journal}{J. Cryst. Growth}}
  \textbf{\bibinfo{volume}{467}}, \bibinfo{pages}{38--46}
  (\bibinfo{year}{2017}).
\newblock
  \urlprefix\url{https://www.sciencedirect.com/science/article/pii/S0022024817301677}.

\bibitem{Repicky_Science2021}
\bibinfo{author}{Repicky, J.} \textit{et~al.}
\newblock \bibinfo{title}{Atomic-scale visualization of topological spin
  textures in the chiral magnet {MnGe}}.
\newblock \textit{\bibinfo{journal}{Science}} \textbf{\bibinfo{volume}{374}},
  \bibinfo{pages}{1484--1487} (\bibinfo{year}{2021}).
\newblock
  \urlprefix\url{https://www.science.org/doi/abs/10.1126/science.abd9225}.

\bibitem{PhysRevLett.102.186602}
\bibinfo{author}{Neubauer, A.} \textit{et~al.}
\newblock \bibinfo{title}{Topological {H}all effect in the {A} phase of
  {MnSi}}.
\newblock \textit{\bibinfo{journal}{Phys. Rev. Lett.}}
  \textbf{\bibinfo{volume}{102}}, \bibinfo{pages}{186602}
  (\bibinfo{year}{2009}).
\newblock
  \urlprefix\url{https://link.aps.org/doi/10.1103/PhysRevLett.102.186602}.

\bibitem{Matsuno_sciadv}
\bibinfo{author}{Matsuno, J.} \textit{et~al.}
\newblock \bibinfo{title}{Interface-driven topological {H}all effect in
  {SrRuO}$_3$-{SrIrO}$_3$ bilayer}.
\newblock \textit{\bibinfo{journal}{Sci. Adv.}} \textbf{\bibinfo{volume}{2}},
  \bibinfo{pages}{e1600304} (\bibinfo{year}{2016}).
\newblock
  \urlprefix\url{https://www.science.org/doi/abs/10.1126/sciadv.1600304}.

\bibitem{Wang2018}
\bibinfo{author}{Wang, L.} \textit{et~al.}
\newblock \bibinfo{title}{Ferroelectrically tunable magnetic skyrmions in
  ultrathin oxide heterostructures}.
\newblock \textit{\bibinfo{journal}{Nat. Mater.}}
  \textbf{\bibinfo{volume}{17}}, \bibinfo{pages}{1087–1094}
  (\bibinfo{year}{2018}).
\newblock \urlprefix\url{https://doi.org/10.1038/s41563-018-0204-4}.

\bibitem{Vistoli2019}
\bibinfo{author}{Vistoli, L.} \textit{et~al.}
\newblock \bibinfo{title}{Giant topological {H}all effect in correlated oxide
  thin films}.
\newblock \textit{\bibinfo{journal}{Nat. Phys.}} \textbf{\bibinfo{volume}{15}},
  \bibinfo{pages}{67–72} (\bibinfo{year}{2019}).
\newblock \urlprefix\url{https://doi.org/10.1038/s41567-018-0307-5}.

\bibitem{Pacheco_NatComm2017}
\bibinfo{author}{Fern{\'a}ndez-Pacheco, A.}, \bibinfo{author}{Streubel, R.},
  \bibinfo{author}{Fruchart, O.}, \bibinfo{author}{Hertel, R.},
  \bibinfo{author}{Fischer, P.} \& \bibinfo{author}{Cowburn, R.~P.}
\newblock \bibinfo{title}{Three-dimensional nanomagnetism}.
\newblock \textit{\bibinfo{journal}{Nat. Commun.}}
  \textbf{\bibinfo{volume}{8}}, \bibinfo{pages}{15756} (\bibinfo{year}{2017}).
\newblock \urlprefix\url{https://doi.org/10.1038/ncomms15756}.

\bibitem{Gao_Nature2020}
\bibinfo{author}{Gao, S.} \textit{et~al.}
\newblock \bibinfo{title}{Fractional antiferromagnetic skyrmion lattice induced
  by anisotropic couplings}.
\newblock \textit{\bibinfo{journal}{Nature}} \textbf{\bibinfo{volume}{586}},
  \bibinfo{pages}{37–41} (\bibinfo{year}{2020}).
\newblock \urlprefix\url{https://doi.org/10.1038/s41586-020-2716-8}.

\bibitem{Mohanta_CommPhys2021}
\bibinfo{author}{Mohanta, N.}, \bibinfo{author}{Okamoto, S.} \&
  \bibinfo{author}{Dagotto, E.}
\newblock \bibinfo{title}{Skyrmion control of {M}ajorana states in planar
  {J}osephson junctions}.
\newblock \textit{\bibinfo{journal}{Comm. Phys.}} \textbf{\bibinfo{volume}{4}},
  \bibinfo{pages}{163} (\bibinfo{year}{2021}).
\newblock \urlprefix\url{https://doi.org/10.1038/s42005-021-00666-5}.

\bibitem{Mascot_NPJQM2021}
\bibinfo{author}{Mascot, E.}, \bibinfo{author}{Bedow, J.},
  \bibinfo{author}{Graham, M.}, \bibinfo{author}{Rachel, S.} \&
  \bibinfo{author}{Morr, D.~K.}
\newblock \bibinfo{title}{Topological superconductivity in skyrmion lattices}.
\newblock \textit{\bibinfo{journal}{npj Quantum Mater.}}
  \textbf{\bibinfo{volume}{6}}, \bibinfo{pages}{6} (\bibinfo{year}{2021}).
\newblock \urlprefix\url{https://doi.org/10.1038/s41535-020-00299-x}.

\end{thebibliography}

\noindent {\bf {\small \\Acknowledgements\\}}
This work was supported by the U.S. Department of Energy, Office of Science, Basic Energy Sciences, Materials Sciences and Engineering Division.

\noindent {\bf {\small \\ Author contributions\\}} All authors contributed to all aspects of this work.

\noindent {\bf {\small \\ Competing interests\\}} The authors declare no competing interest.

\noindent {\bf {\small \\ Additional information\\}}  
Correspondence should be addressed to N. Mohanta.

\end{document}